\documentclass{PoS}
\usepackage{epsfig}

\PoS{PoS(BDMH2004)098}

\title{New Gamma-Ray Probe of the Baryonic Dark Matter}

\author{\speaker{Anatoli Iyudin}\\
        MPE, Garching, Germany, and Moscow State Univ., Skobeltsyn Insitute of Nuclear Physics, Moscow, Russia\\
        E-mail: \email{ani@mpe.mpg.de, aiyudin@srd.sinp.msu.ru}}

\author{Vadim Burwitz\\
        Max-Planck-Institut f\"ur extraterrestrische
 Physik, Postfach 1312, 85741 Garching, Germany\\
        E-mail: \email{burwitz@mpe.mpg.de}}
\author{Jochen Greiner\\
        Max-Planck-Institut f\"ur extraterrestrische
 Physik, Postfach 1312, 85741 Garching, Germany\\
        E-mail: \email{jcg@mpe.mpg.de}}
\author{Guido DiCocco\\
        IASF-Bologna, Bologna, Italy\\
        E-mail: \email{dicocco@bo.iasf.cnr.it}}
\author{Stefan Larsson\\
        Stockholm University, Stockholm, Sweden\\
        E-mail: \email{stefan@astro.su.se}}

\abstract{We expand on the recently reported detections of the gamma-ray resonant 
absorption along 
the line of sight toward gamma-ray bright quasars (QSOs), like 3C279. We propose to use this novel gamma-ray absorption
 method to study
the Dark Matter distribution in the Milky Way, as well as in the Local Group. 
\par
Properties of the absorber that was detected on the sight lines towards 
gamma-ray bright QSOs at zero redshift are discussed. We compare our 
results with the expected Dark Matter distributions in the halo of Milky Way, 
that were simulated in boundaries of different CDM models.
\par
Application of this new method to study
evolution of CDM in the QSO host galaxies, and of baryons distribution in the halo of galaxies in the Local Universe is proposed.
}

\FullConference{Baryons in Dark Matter Halos\\
		 5-9 October 2004\\
		 Novigrad, Croatia}

\ShortTitle{Baryonic DM}

\begin{document}

\section{Introduction}

One of the pressing problems in the modern cosmology is that the nature of the
dark matter is at present unknown. The properties of dark matter on
large scales are compatible with a heavy, collision-less species which
clusters gravitationally, cold dark matter (CDM). But
there are inconsistencies between simulations and observations of the dark matter halo of
galaxies in boundaries of CDM models. While the outer slope derived from the numerical simulations [9], [10], [12], [18], and [19] is consistent with
current observations, there are problems with fitting the inner slope of
halos.
Observations of rotation curves of many dwarf galaxies, indicating
almost constant density cores, suggest that the inner density profiles
in these systems are much shallower than found in simulations
[4], [5], and [22].

We propose to use a new method, namely, a recently introduced $\gamma$-ray probe of the absorbing columns on the line sights towards bright quasars (QSOs) [14] to solve the discrepancy between observations and simulations of the halo density profiles.
The $\gamma$-ray absorption can be used to probe higher column densities than those accessible for the longer wavelengths. Additionally, it is not sensistive to the absorber ionization or chemical state.
\par
In our method we rely on the photoabsorption processes that happen on the nuclei. Apart of the well known high-energy photons attenuation processes, like Compton scattering and pair production, with a rather smooth energy dependence of the cross section at E$_{\gamma}$$\geq$100 keV [13], there are three more photoabsorption processes to mention [14].
Namely, the photoabsorption cross-section on nuclei have three resonant-like peaks in the cross section, at energies of $\sim$7 MeV (``pygmy'' dipole resonance (PDR)), 20-30 MeV (giant dipole resonance (GDR)), and $\sim$325 MeV (${\Delta}$-resonance) [1].
The best studied of these three processes are GDR, and $\Delta$-isobar resonance [1].
\par
From the ratio of the typical absorption cross sections in the $\gamma$-ray and X-ray regimes follows that $\gamma$-rays probe the column densities of the order of $\sim$10$^{26}$ cm$^{-2}$ via the resonant $\gamma$-absorption, while in X-ray regime we probe column densities of the order of $\leq$10$^{25}$ cm$^{-2}$.

\section{Absorbing columns near Galactic Center}
Results used to constrain the shape of the baryonic matter distribution in the region of Galactic Center (GC) are based on the data acquired by the gamma-ray telescopes COMPTEL and EGRET on-board of the Compton Gamma-Ray Observatory (CGRO).
A detailed description of the instruments and data analysis is given in [21] for COMPTEL, and in [23], and [16] for EGRET.
\par
We analysed COMPTEL data using the maximum-likelihood method (SRCFIX), that evolved from the work on diffuse emission modeling [24]. The analysis of EGRET data is also based on the maximum likelihood analysis of the observed region [16].
\par
In this paper we will mostly use results of EGRET, that has best sensitivity near the 300 MeV, i.e near the energy of the ${\Delta}$-resonance absorption for systems with a small redshift.
\begin{figure}
  \centering
\includegraphics[bb=105 565 495 699,width=14.5cm,clip]{EUID-spectrum_DM-cusp.ps}
  \caption{{\it Left}: A fit to the 3EG J1744-3011 time-averaged SED, that includes the ${\Delta}$ resonance photon absorption (red line) in the matter of Milky Way. Black line shows ``Band''-function fit. {\it Right}: The comparison of the DM density profiles derived from the numerical simulations, adapted from [6], with the preliminary normalized values of the baryonic absorption columns derived from the amplitude of the ${\Delta}$ resonance photon absorption troughs in the SEDs of EUIDs listed in the Table 1. Here ``can'' stands for canonical distribution [2]; ``PS'' stands for distribution of [20]; ``M\&'' stands for [17], and ``NFW'' is from [18].}
\label{fig1}
\end{figure}
The differential photon flux from AGN at the Earth can be written as a  function of the photon energy and of redshift:
$\frac{dN}{dE}= ( \frac{dN}{dE} )_{unabsorbed}\cdot e^{-\tau(E,z)} \hspace{2mm}$.
\par
The dependence of $\tau$ on $E$ and $z$ is quite complex,
to simplify it we assume that we are dealing with two absorbers,
one in the QSO host galaxy, and the second absorber in the Milky Way.
\par
As a first step of a SED analysis, we fit a smooth function to the spectrum, which we choose to be the so-called ``Band''-function [3], that describes well not only power-law spectra, but also spectra with a break. In the second fit we use as a fit function the sum of the ``Band''-function and a gaussian.
The significance of the fit improvement we evaluate using the probability for the spurious improvement of the fit from the value of ${\Delta}{\chi}^2$=${\chi}^2_{band}$-${\chi}^2_{band+gauss}$ for 3 d.o.f. as a test for such an improvement, similar to [8].
Some of the $\gamma$-ray bright sources in the GC region, can be used to probe the baryonic matter distribution in the center of Milky Way. The very first results of such a ``probe'' are discussed below.
\begin{table*}[hbt]

\begin{tabular*}{147mm}{||c||c||c||c||c||c||c||c||@{\extracolsep\fill}}
\hline
 Object  & $l^o$ & $d^o$ & ${\theta}_{95}$, deg. &
${\Delta}d$, (pc) & $N_H$$\times$ $10^{26}$ & ${N_H}^*$  &No.  \\
\hline
\hline
 3EG J1746-2851 & 00.11 & -0.04  & 0.13 & 17.366 & $24^{+6}_{-10}$ & $6.5^{+1.6}_{-2.7}$ & 1    \\
\hline
 3EG J1744-3011 & -1.15 & -0.52  & 0.32 & 187.26 & $35^{+29}_{-23}$ & $9.5^{+7.8}_{-6.2}$ & 2    \\
\hline
 3EG J1736-2908 & -1.21 & 1.56  & 0.62 & 275.69 & 4.1$\pm$2.4 & 1.11$\pm$0.65  & 3   \\
\hline
 3EG J1741-2312 & 04.42 & 3.76  & 0.57 & 813.02 & 3.8$\pm$2.5 & 1.02$\pm$0.68  & 5   \\
\hline
 3EG J1717-2737 & -2.33 & 5.95  & 0.64 & 895.92 & 3.1$\pm$2.2 & 0.84$\pm$0.59   & 6  \\
\hline
 3EG J1741-2050 & 06.44 & 5.00  & 0.63 & 1146.1 & 3.9$\pm$2.4 & 1.07$\pm$0.65   & 7  \\
\hline
\end{tabular*}
\par
$^*$ Value normalized to the mean value of $N_H$=(3.70$\pm$1.19)$\times$10$^{26}$ $cm^{-2}$ for ${\Delta}$d$>$275~pc.
\caption{\bf EUIDs, their projected distance from the GC, and absorbing columns for these projected distances.}
\label{tab1}
\end{table*}
We have used SEDs of 6 EGRET Unidentified sources (EUIDs) [11] to derive baryonic absorbing columns near GC (Table 1). 3EG J1756-2851 is at a redshift of $\sim$0.85 [14]. By analogy with 3EG J1736-2908 which was identified with the Seyfert I galaxy GRS 1734-292 [7], and [15], we consider 4 other EUIDs also as extragalactic sources. For 5 out of 6 EUIDs the fit and the derived absorbing column was made in the region of ${\Delta}$-resonance (Fig. 1, left). Only for 3EG J1746-2851 we have used results of the fit at the energies of GDR, e.g around 25 MeV.
In this case we derived the lowest possible value of the absorbing column. Four columns derived for EUIDs at the projected distances of ${\Delta}$d$>$275~pc were used to calculated weighted mean value of the absorbing column, that was normalized to the calculated DM density profiles of Fig. 1 (right) at $r$=1~kpc.
Figure 1 (right) shows the Milky Way central core/cusp profiles derived by the numerical simulations that are (rather crudely) compared with the baryonic absorbing columns measured at different projected distances from the GC.
\par
The incentive of our paper was to demonstrate the potential of the $\gamma$-ray absorption method for this or similar studies. Still, already these results give a hint of a more shallow slope of the inner cusp than that from the numerical simulations [17], or [18]. It is clear that more sensitive measurements of EUID SEDs in the GC region are needed to provide definite conclusion on the preferable baryonic matter profile close to the Galactic Center. Such measurements can be performed by the forthcoming new gamma-ray mission {\it GLAST}, that will be launched in 2007, see http://www-glast.slac.stanford.edu, for the {\it GLAST} description.


\begin{thebibliography}{99}

\bibitem{ahrens85}
Ahrens, J., {\it Nucl.Phys}\ {\bf A446}, 229c (1985).
\bibitem{bahcall80}
Bahcall, J.N., and Soneira, R.M., {\it ApJS}\ {\bf 44}, 73 (1980).
\bibitem{band93}
Band, D., Matteson, J., Ford, L., et al., {\it ApJ}\ {\bf 413}, 281 (1993).
\bibitem{de Blok et al. 2001} de
Blok, W.J.G., McGaugh, S.S., Bosma, A. and Rubin, V.C., {\it ApJ}\
{\bf 552}, L23 (2001).
\bibitem{van den Bosch et al. 2000}
van den Bosch, F.C., Robertson, B.E.,
Dalcanton, J.J. and de Blok, W.J.G., {\it AJ}\ {\bf 119}, 1579 (2000).
\bibitem{Buckley et al. 2002}
Buckley, J., Burnett, T., Sinnis, G., et al.,
%``Gamma-Ray Summary Report''
[astro-ph/0201160] (2002).
\bibitem{Di Cocco et al. 2004}
Di Cocco, G., Foschini, L., Grandi, P., et al.,
%``INTEGRAL observation of 3EG J1736-2908''
[astro-ph/0406300] (2004).
\bibitem{freeman99}Freeman, P.E., Graziani, C., Lamb, D.Q., et al., {\it ApJ}\ {\bf 524}, 753 (1999).
\bibitem{Fukushige et al. 2004}
Fukushige, T., Kawai, A. and Makino, J.,
%``Structure of Dark Matter Halos From Hierarchical Clustering. III. %Shallowing
%of The Inner Cusp,''
{\it ApJ}\ {\bf 606}, 625 (2004).
\bibitem{Ghigna et al. 2000}
Ghigna, S., Moore, B., Governato, F., Lake, G., Quinn, T. and Stadel, J.,
%``Density profiles and substructure of dark matter halos: converging %results at
%ultra-high numerical resolution,''
{\it ApJ}\ {\bf 544}, 616 (2000).
\bibitem{hartman99}Hartman, R.C., Bertch, D.L., Bloom, S.D., et al., {\it ApJS}\ {\bf 123}, 79 (1999).
\bibitem{Hayashi et al. 2003}
Hayashi, E., et al.,
%``The Inner Structure of LCDM Halos II: Halo Mass Profiles and LSB %Rotation Curves,''
astro-ph/0310576 (2003).
\bibitem{hubbell71}Hubbell, J.H., {\it Atomic Data}\ {\bf 3}, 241 (1971).
\bibitem{iyudin04} Iyudin, A.F., et al., {\it A\&A}\ , submitted (2004).
\bibitem{marti98} Mart\'{\i}, J., Mirabel, I.F., Chaty, S., Rodr\'{\i}guez, L.F., {\it A\&A}\ {\bf 330}, 72 (1998).
\bibitem{mattox96}Mattox, J.R., Bertsch, D.L., Chiang, J., et al., {\it ApJ}\ {\bf 461}, 396 (1996).
\bibitem{moore98}Moore, B., Governato, F., Quinn, T., Stadel, J., and Lake, G., {\it ApJ}\ {\bf 499}, L5 (1998).
\bibitem{Navarro et al. 1997}
Navarro, J.F., Frenk, C.S. and White, S.D.M.,
%``A Universal density profile from hierarchical clustering,''
{\it ApJ}\ {\bf 490}, 493 (1997).
\bibitem{Navarro et al. 2004}
Navarro, J.F., et al.,
%``The Inner Structure of LambdaCDM Halos III: Universality and Asymptotic
%Slopes,''
{\it MNRAS}\ {\bf 349}, 1039 (2004).
\bibitem{Persic et al. 1996}
Persic, M., Salucci, P., and Stel, F.,
%``The Universal rotation curve of spiral galaxies - I. The dark matter connection''
{\it MNRAS}\ {\bf 281}, 27 (1996).
\bibitem{schoenfelder93}Sch\"onfelder, V., Aarts, H., Bennett, K., et al., {\it ApJS}\ {\bf 86}, 657 (1993).
\bibitem{Simon et al. 2003}
Simon, J.D., Bolatto, A.D., Leroy, A. and Blitz, L.,
%``Dark Matter in Dwarf Galaxies: Latest Density Profile Results,''
astro-ph/0310193 (2003).
\bibitem{thompson93}Thompson, D.J., Bertsch, D.L., Fichtel, C.E., et al., {\it ApJS}\ {\bf 86}, 629 (1993).
\bibitem{vandijk96}van Dijk, R., {\it PhD Thesis}, Amsterdam Univ. (1996).

\end{thebibliography}
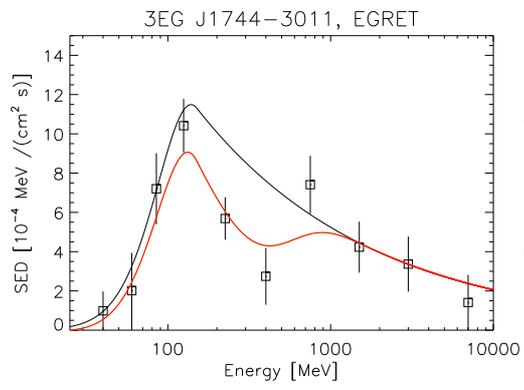
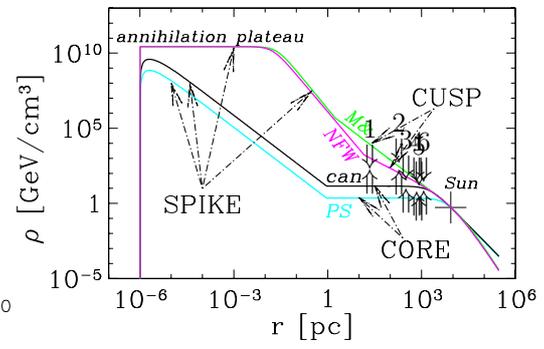
\end{document}